# Implementing Resolute Choice Under Uncertainty

Jean-Yves Jaffray
LIP6, UPMC (Paris 6)
F-75252 Paris cedex 05

## Abstract

The adaptation to situations of sequential choice under uncertainty of decision criteria which deviate from (subjective) expected utility raises the problem of ensuring the selection of a non-dominated strategy. In particular, when following the suggestion of Machina and McClennen of giving up separability (also known as consequentialism), which requires the choice of a substrategy in a subtree to depend only on data relevant to that subtree, one must renounce to the use of dynamic programming, since Bellman's principle is no longer valid. An interpretation of McClennen's resolute choice, based on cooperation between the successive Selves of the decision maker, is proposed. Implementations of resolute choice which prevent Money Pumps, negative prices of information or, more generally, choices of dominated strategies, while remaining computationally tractable, are proposed.

## 1 INTRODUCTION

From the point of view of decision aiding in dynamic choice situations, subjective expected utility (SEU) theory possesses two appealing features: (i) it guarantees rational choices, in the sense that a SEU maximizing decision maker (DM) is immune to the manipulations known as Dutch Books or Money Pumps, never displays any aversion to information and, more generally always selects a "non-dominated" strategy); (ii) optimal strategies can be determined by backward induction (i.e., by "rolling back" the decision tree) due to the validity of Bellman's principle of dynamic programming.

The use of SEU theory requires the assessment of two characteristics: (i) the von Neumann-Morgenstern (vNM) utility function of the DM on the consequences, and (ii) his subjective probability distribution on the events. Standard utility and probability assessment methods exist. They may not work satisfactorily, and yield incoherent evaluations, for the following reason: while SEU theory is normatively irreproachable, it performs extremely poorly as a descriptive model, as repeatedly shown by numerous experiments (Camerer and Weber 1992).Thus choices made by the DM are likely to exhibit patterns which cannot be accounted for by the model and make any probability or utility assesment questionable.

One way out of this difficulty is to try and identify "biases" of the DM w.r.t. SEU theory and modelize a "debiased" behaviour. Indeed, decision aiding consists in part in helping the DM to avoid irrational choices based on psychological as well as on perceptual misrepresentations. However the DM may feel strongly for his spontaneous judgments, and thus be reluctant to abide by the decisions prescribed by the model.

Another way out, which has been largely explored during the last decade, is to turn to different, more flexible, decision models, which in particular are consistent with the most robust choice biases. Some of these models use a probabilistic representation, but evaluate decisions according to criteria which are no longer linear in probabilities. Some other have recourse to non-probabilistic representations of uncertainty, such as upper/lower probabilities, belief functions and possibility/necessity functions.

However models departing from SEU raise new difficulties. Even if the DM keeps the same criterion throughout time (and just updates his probabilities, beliefs, etc... when receiving information) his preference pattern is no longer dynamically consistent, in the sense that future decisions judged optimal now may no longer be still considered to be so when future comes. As a consequence, if dynamic programming is nonetheless used to select a strategy (game theory has strong arguments in favor of that solution), the strategy chosen may well be dominated, i.e., there may exist another feasible strategy yielding a better consequence, whatever events occur.

Since there is no such difficulty with one-shot decision problems, why not transform dynamic decision problems into static ones, by choosing at the initial period a strategy which is judged optimal at the moment, and never revising it later ? This solution has the obvious drawback of requiring the direct comparison of all feasible strategies and may be computationally intractable even in medium-size decision problems. Moreover, the question of its psychological feasibility arises, since it requires the DM to abide at later periods by decisions he may no longer consider optimal.

However, other forms of collaboration between the successive Selves of the DM beside the dictatorship of the initial Self can be contemplated. Following (McClennen 1990) we call *resolute choice* the accomplishment of any such collaboration process. Decision aiding, requires explicit methods for implementing resolute choice solutions. Tractability considerations show that it would prove difficult to dispense with the advantages of dynamic programming ; on the other hand, the attainment of resolute choices implies the interdependence of all local choices, hence the presence of nonseparability making Bellman's principle invalid. We shall see that it is in fact possible to get round this apparent contradiction.

## 2 SEU THEORY AND THE NEED FOR ALTERNATIVE THEORIES

### 2.1 SEU THEORY

Subjective Expected Utility (SEU) theory evaluates a decision d giving gain $c_i$ if event $A_i$ obtains by

$$U(d) = \sum_{i=1}^{n} P(A_i) u(c_i) , \qquad (1)$$

where probability P, defined on the set of events, and utility u(.), defined on the set of consequences (gains), are subjective, i.e., depend on the DM.

### 2.2 ITS FAILURE TO EXPLAIN THE ELLSBERG EXPERIMENT

An urn contains 90 balls, out of which 30 are R(ed) and 60 are B(lack) or Y(ellow). A ball is to be drawn at random and the subjects are asked about their choice when facing alternatives $d_R$ and $d_B$ and when facing $d_{R \cup Y}$ and $d_{B \cup Y}$ (see Table 1).

Table 1

| ball | R | B | Y |
|---|---|---|---|
| $d_R$ | 100 | 0 | 0 |
| $d_B$ | 0 | 100 | 0 |
| $d_{R \cup Y}$ | 100 | 0 | 100 |
| $d_{B \cup Y}$ | 0 | 100 | 100 |

Ellsberg has observed the predominant choice pattern : $d_R$ (against $d_B$) ; $d_{B \cup Y}$ (against $d_{R \cup Y}$). These choices are not explainable by SEU (nor, in fact by any theory involving subjective probabilities) since they would imply P(R) > P(B) and P(R∪Y) < P(B∪Y), in contradiction with additivity.

On the other hand, this choice pattern can be accounted for by the Schmeidler (1989) model, in which non-additive weights ponder the utilities in relation with their ranking ; more precisely, a decision d, which gives a gain $c_i$ if event $A_i$ obtains, with indices i = 1,...,n chosen such that $c_i \leq c_{i+1}$, has value

$$V(d) = u(c_1) + \sum_{i=2}^{n} \Pi (\bigcup_{j=i}^{n} A_j) [u(c_i) - u(c_{i-1})] . \quad (2)$$

V is called a Choquet Expected Utility (CEU) criterion. Mathematically speaking, Π is a capacity and the right handside of (2) is a Choquet integral.

The observed choices are consistent with Π(R) > Π(B) and Π(R∪B) < Π(B∪Y), which can be explained psychologically by *ambiguity aversion* : a gain on Y, with an ambiguous probability located in [0,2/3] is less attractive than the same gain with probability 1/3 ; similarly, a gain on R∪Y, with probability in [1/3,1], is less attractive than the probability 2/3 gain on B∪Y.

Thus CEU is an interesting generalization of SEU from the descriptive point of view since it can take into account subjects' attitude with respect to ambiguity as well as other psychological traits such as the certainty effect (Kahnemann and Tversky, 1979). There is clearly a price to pay for this flexibility from the prescriptive (decision aiding) point of view, since the evaluation of Π, a non-additive set function, is likely to require a much greater effort than the evaluation of a probability (which is completely determined by its values on singletons). Nonetheless, the balance would generally be considered to turn in favor of the use of CEU theory rather than of SEU theory were it not for the appearance of serious problems in trying to extend the use of CEU theory or other "non-EU" models to dynamic decision making situations. The seminal paper on dynamic choice is (Hammond 1988).

## 3 DYNAMIC DECISION MAKING AND SOME BASIC CONCEPTS

As opposed to static (one-shot) decision problems, dynamic decision problems are concerned with multiple, sequential and conditional, decisions. Such a problem is conveniently represented (in the finite case) by a decision tree, in which the DM's task is to choose a strategy (also called a plan), i.e., a sequence of conditional decisions.

Given a strategy and the true elementary event, the root-to-leaf path describing the succession of decisions taken and events observed is completely determined, and so is the gain. Thus, a strategy defines a mapping *elementary event → gain* , which can be evaluated by the DM's criterion , say, a CEU criterion ; moreover, a substrategy, i.e., the trace of a strategy on the subtree rooted at some decision node, defines a similar mapping (restricted to the elementary events consistent with the information) and can also be evaluated by the DM's local criterion (which may, or may not, be the same CEU criterion). An *optimal* substrategy at a decision node is a substrategy which has the highest value among all feasible substrategies at that node.

Note that at any decision node the DM is only able to enforce the first decision of any strategy he might contemplate : the following decisions will be taken by his future Selves. When, unlike the *myopic* DM in (Strotz 1956), is aware of this fact, he will make a distinction between feasible and available substrategies. In particular, a straightforward approach (but by no means the only one) for solving a dynamic problem consists in using backward induction and progressively evaluating the still available substrategies at the various decision nodes. Let us try this solution on an example (derived from the Ellsberg experiment).

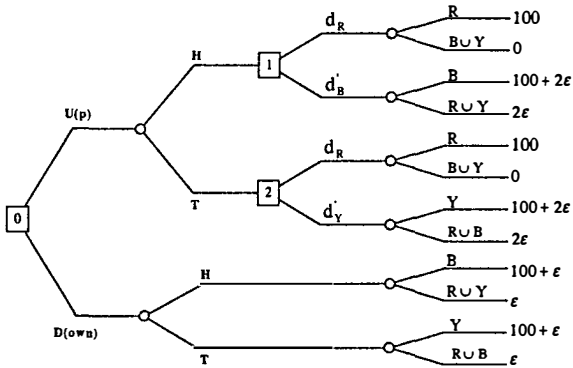

Figure 1

Example 1

Consider again an Ellsberg urn, and events R(ed), B(lack) and Y(ellow), and moreover a coin and equiprobable events H(eads) and T(ails). Event$(H \cap B) \cup (T \cap Y)$ has probability 1/3. Figure 1 displays the decision problem. Suppose that the DM is a CEU maximizer, with $\Pi(R) = 1/3 > \Pi(B) = \Pi(Y)$ and $u = Id$.

Let $V_i$ be his criterion at node $i$. At node 1, $V_1(d_R) = 100 \, \Pi(R)$ and $V_1(d'_B) = 2\varepsilon + 100 \, \Pi(B)$, hence, for $\varepsilon > 0$ small enough, $d_R$ is chosen ; for the same $\varepsilon$, $V_2(d_R) > V_2(d'_Y)$ and $d_R$ is also chosen at node 2.

Using backward induction, the DM then compares at node 0 strategies (U,$d_R$ if H, $d_R$ if T) and D and prefers the latter since

$V_0(U,d_R \text{ if } H, d_R \text{ if } T) = 100/3 < \varepsilon + 100/3 =$
$\varepsilon + 100 \, \Pi((H \cap B) \cup (T \cap Y))$ .

Thus D is finally chosen, despite the fact that there exists a feasible strategy (U,$d'_B$ if H, $d'_Y$ if T) which (*strictly*) *dominates* it, i.e., would bring to the DM a gain superior by $\varepsilon$ to the gain offered by D, whatever happens. ∎

The recursive construction of one's strategy by anticipating ones' future choices is known as *sophisticated choice* (Strotz, 1956). The preceding example shows that, when combined with CEU maximization at the various decision nodes, a sophisticated DM can end up chosing a dominated strategy.

This example clearly exploits the following fact:whereas the most appealing strategy for the DM at node 0 is (U,$d'_B$ if H, $d'_Y$ if T) he considers it as *de facto* infeasible, since he foresees that, would he choose initially U , his preferences at node 1 and node 2 would make him continue differently and actually play strategy (U, $d_R$ if H, $d_R$ if T). For this reason, he settles for the less attractive but enforceable strategy D. Thus his suboptimal choices are the result of his dynamic inconsistencies in the following sense : a DM is *dynamically consistent* (DC) when the DM considers it optimal at every node to continue as planned by the original optimal strategy.

The possibility of decomposing the SEU criterion as

$$\sum_i P(e_i) u(c_i) = \sum_j P(A_j) [ \sum_{e_i \subset A_j} P(e_i/A_j) u(c_i) ] \quad (3)$$

shows immediately that SEU maximizers are DC provided they update probabilities according to Bayes' rule. Note that the criterion at node j, where event $A_j$ is known to obtain, takes only into account the gains on the subevents of $A_j$ ; what happens outside $A_j$ is irrelevant.

According to a general definition, *separability* (SEP) holds when the DM's optimal strategy in any subtree does not depend on the rest of tree. Thus, DC and SEP hold true for the SEU criterion. Together DC and SEP imply the validity of Bellman's principle of dynamic programming and thus the possibility of generating an optimal strategy by backward induction.
On the other hand, when applying a CEU criterion at each node, SEP is satisfied and DC is not. As we shall see now , it is also possible to conceive non-EU behavior which satisfies DC but not SEP and offers some guarantees from the normative point of view.

# 4 RESOLUTE CHOICE AND ITS IMPLEMENTATION

## 4.1 RESOLUTE CHOICE

The thorough analysis of dynamic choice under uncertainty in (McClennen 1990) resulted in his proposal of an alternative to SEU theory, the so-called *resolute choice* solution, which we shall introduce through our example.

Example 1 (continued)

The DM can always achieve an undominated strategy by enforcing the strategy he judges optimal at node 0 and thus being DC. This is strategy (U, d'$_B$ if H, d'$_Y$ if T), which means that at node 1 he accepts to choose d'$_B$ against d$_R$.

Consider now the modified tree where, at node 2 d'$_Y$ has been replaced by d'$_B$ and D(own) yields 0 whatever happens. The DM being ambiguity averse, it is likely that
$1/3 = \Pi(R) > \Pi((H \cap R) \cup (T \cap Y)) =$
$\Pi((H \cap B) \cup (T \cap R)) > \Pi(B)$,
in which case, the best strategy at node 0 is (U, d$_R$ if H, d$_R$ if T), Now, its enforcement requires that, at node 1, the DM accept to choose d$_R$ against d'$_B$. This shows that choice at node 1 depends on data outside the subtree rooted at this node. ∎

Thus, a DM can depart from SEU maximization, and still make non-dominated choices, by being dynamically consistent. There is however a price to pay, which is giving up SEP (Machina 1989).

Submission of later choices to initial preference is of course only a special case of more general forms of compromise between present and future wishes. According to (McClennen 1990), *"the theory of resolute choice is predicated on the notion that the single agent who is faced with making decisions over time can achieve a cooperative arrangement between his present self and his relevant future selves that satisfies the principle of intrapersonal optimality"*.

There is fundamental problem with resolute choice which is the question of its psychological feasibility. Why should the Selves cooperate ? As a matter of fact, sophisticated choice is the noncooperative solution proposed by game theory : as a subgame perfect Nash equilibrium (Rasmusen 1989 ; Karni and Safra 1989), it constitutes a credible solution. . On the other hand, it can be argued that, being the successive Selves of the same DM, the players: (i) have part of their interests in common; and (ii) cannot hide their intentions from one another, which should facilitate their cooperation.

We shall admit that the Selves have consensus goals and are able to cooperate, at least to some extent. Specifically, we endow each Self with a choice criterion which is not influenced by what happens outside his subtree (thus in the absence of cooperation SEP would hold).

The willingness of the Selves to cooperate shall be expressed either by commitments to rule out certain decisions or by (possibly limited) obedience to earlier Selves' recommendations.

The main consensus goal we shall consider is the choice of an *undominated strategy* : a strategy in a decision tree is *dominated* when there exists another feasible strategy yielding a superior gain whatever happens and a strictly superior gain for some elementary event.

A less demanding goal is *immunity to Money Pumps* :
a DM is the victim of a Money Pump when, at some decision node, he accepts to pay a fee and find oneself in an earlier decision position again (more precisely, he gets to the root of a decision subtree which is identical to a previously faced one). Another less demanding goal is *nonnegative price of information* : information received before, rather than after, decision making enlarges the set of feasible strategies, since the DM is free to make, or not, his later decisions contingent on these data. Thus the DM should never accept to pay for not receiving information. Strategies involving either a Mony Pump or a negative price of some information are clearly instances of dominated strategies.

We shall successively consider various forms of commitment of the Selves and examine wether or not they permit to achieve these consensus goal. Note that any amounf of cooperation makes SEP invalid, which in turn implies that dynamic programming will, at best, only be useful as an auxiliary tool.

## 4.2 JUSTIFIABLE CHOICE

Any form of cooperation ensuring the final selection of an undominated strategy must in particular ensure that no Self ever chooses a decision which is not part of at least one undominated strategy. *Justifiable choice* is the particular instance of resolute choice in which each Self exactly makes this minimal commitment. Let us (i) see how it can be implemented and (ii) examine what can be achieved by it.

Jointly, these commitments amount to replacing the decision tree T by the "subtree" T$_0$ which is composed of all the paths which could result from at least one undominated strategy. It is not easy, in general to construct T$_0$. In practice, one would only generate part of it through a limited number of undominated strategies, those maximizing linear forms (expected gains)

$\sum_i \alpha_i c_i$ for various sets of positive weights

(probabilities) ($\alpha_i$) on the elementary events e$_i$ (each strategy being standardly determined by dynamic programming, which is, of course, valid for expectations) and restrict oneself to the "subtree" $T_1$ spanned by these strategies (and the chance moves). We call *justifiable strategies* the strategies in $T_0$. They are indeed justifiable *ex post*, since there is always an undominated strategy which, for the nested sequence of true events, would have lead to the same sequence of decisions. Obviously, for different sequences of true events, the decisions resulting from a justifiable strategy coincide with those decisions resulting from another undominated strategy. Thus, even a justifiable strategy of $T_1$ cannot be associated with particular weighting systems $(\alpha_i)$ and may well be a dominated strategy.

In the absence of further cooperation, the strategy finally selected should be the sophisticated strategy in $T_0$ (in practice in $T_1$), which can be determined by backward induction (dynamic programming): each Self, in turn, anticipates later Selves' choices and optimizes his own choice in $T_0$ (commitment) according to his non-EU criterion. Note that the introduction of subtree $T_0$ has allowed one to recover SEP.

As for the rationality properties of justifiable choice, it is straightforward that the DM is immune to Money Pumps: the edge (decision) "accept to pay" cannot be part of any undominated strategy in T ("refuse to pay" offers the same opportunities plus the fee), hence this edge is not part of $T_0$ either. For similar reasons, this DM will never attribute a negative price to information.

However, a justifiable strategy can be dominated as shown by the following example :

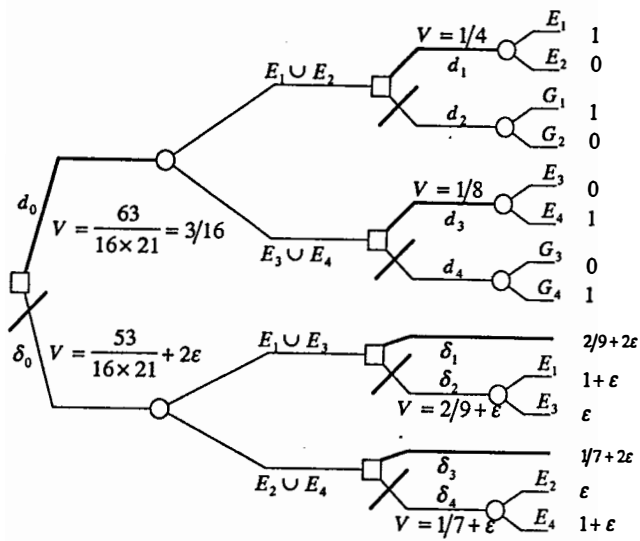

Figure 2

Example 2

Elementary events: $e_i$, $i = 1,...,8$. $E_1 = e_1 \cup e_2$; $E_2 = e_3 \cup e_4$; $E_3 = e_5 \cup e_6$; $E_4 = e_7 \cup e_8$; $G_1 = e_1 \cup e_3$; $G_2 = e_2 \cup e_4$; $G_3 = e_5 \cup e_7$; $G_4 = e_6 \cup e_8$; hence, $E_1 \cup E_2 = G_1 \cup G_2$ and $E_3 \cup E_4 = G_3 \cup G_4$.

Information on the events is characterized by lower probability $\Pi$, which is a belief function with Möbius masses (basic probability assignment) :

$\varphi(E_1 \cup E_2) = 1/4$; $\varphi(E_3 \cup E_4) = 3/8$; $\varphi(E_1) = \varphi(E_2) = 1/8$; $\varphi(E_3) = \varphi(E_4) = 1/16$.
Thus $\Pi(E_1) = \Pi(E_2) = 1/8$; $\Pi(E_3) = \Pi(E_4) = 1/16$; $\Pi(E_1 \cup E_3) = \Pi(E_1 \cup E_4) = 3/16$.

We assume that the DM updates lower probabilities by the generalized Bayesian rule

$\Pi(A/B) = \Pi(A/[\Pi(A) + 1 - \Pi(A \cup B^c)]$, hence :
$\Pi(E_1 / E_1 \cup E_2) = 1/4$; $\Pi(E_4 / E_3 \cup E_4) = 1/8$;
$\Pi(E_1 / E_1 \cup E_3) = 2/9$; $\Pi(E_4 / E_2 \cup E_4) = 1/7$.

We moreover assume that the DM is a CEU maximizer, with $U = \text{Id}$ and capacity $\Pi(./B)$ when B is the available information. He is also a sophisticated DM. The decision tree T is shown on Figure 2. $\varepsilon$ is an arbitrarily small positive number. T is generated by the justifiable strategies, i.e., $T_0 = T$, since the linear form defined by the $(\alpha_i)$ is maximized by strategy $(d_0, d_1, d_4)$ when $\alpha = (0, 1/4, 0, 1/4, 0, 1/4, 0)$, by strategy $(d_0, d_2, d_3)$ when $\alpha = (1/4, 0, 1/4, 0, 0, 1/4, 0, 1/4)$, by strategy $(\delta_0, \delta_2, \delta_4)$ when $\alpha_i = 1/8$ for all i, and by strategy $(\delta_0, \delta_1, \delta_3)$ when $\alpha = (0, 1/8, 0, 7/8, 15/16, 0, 1/16, 0)$.

The sophisticated strategy is $(d_0, d_1, d_3)$, which is dominated by $(d_0, d_2, d_4)$.    ∎

Thus a stronger form of cooperation between the Selves is required to guarantee the selection of an undominated strategy.

### 4.3   A COOPERATIVE DECISION PROCESS

#### 4.3.1   Unlimited cooperation

The progressive construction of an undominated strategy by a recursive procedure can be easily devised.
Firstly, as in subsection 4.2, a series of weighting systems $(\alpha_i)$, which can be interpreted as probabilities on the elementary events $(e_i)$, are generated. For every system $(\alpha_i)$, a strategy maximizing the corresponding linear form in the decision tree T is determined (by dynamic programming). This strategy is undominated in T. We call it the $(\alpha_i)$-*strategy*. These strategies (together with the chance moves) span a "subtree" $T_1$ of T.

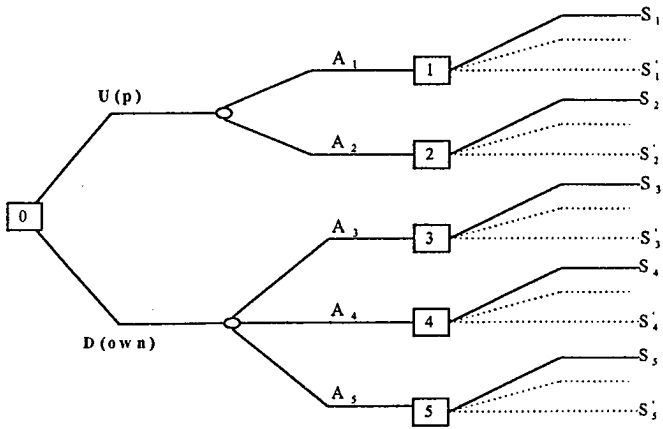

**Figure 3**

We use Figure 3, where a portion of subtree $T_1$ is represented, to explain the principle of the procedure, which rolls back $T_1$.

The recurrence assumption is that undominated substrategies $S_k$ have been tentatively selected at nodes k = 1,...,5, and that each of them is part of (at least) one of the strategies spanning $T_1$ and thus is associated with some weighting system $(\alpha_{ki})$. For instance $S_1$ and (U, $S_1$ if $A_1$, $S'_2$ if $A_2$) might be $(\alpha_{1i})$-substrategies (i.e., substrategies of $(\alpha_{1i})$-strategies), whereas $S_2$ and (U, $S'_1$ if $A_1$, $S_2$ if $A_2$) might be $(\alpha_{2i})$-substrategies, and $S_3$ and (D, $S_3$ if $A_3$, $S'_4$ if $A_4$, $S'_5$ if $A_5$) might be $(\alpha_{3i})$-substrategies, etc... By selecting at node 0 either (U, $S_1$ if $A_1$, $S'_2$ if $A_2$) or one of the other $(\alpha_{ki})$-substrategies, one always selects an undominated $(\alpha_i)$-substrategy. It becomes then possible to select the best of them according to the DM's real criterion at node 0.

Thus we have moved one step back in $T_1$. Clearly, this procedure finally selects an undominated strategy at the root.

Although limited to a small number of substrategies, optimizations at the decision nodes take into account the real preferences of the Selves, so that the strategy finally select must reflect them to some extent.

Yet it seems unrealistic to admit that the Selves are ready to switch from a given decision to any other less attractive decision for ensuring an undominated choice : the willingness to cooperate might have limits.

### 4.3.2 Limited cooperation

The rather vague concept of willingness to cooperate can be given an operational meaning when the Selves have the same cardinal criterion such as CEU with the same utility u(.). One can for instance assume that every Self is willing to accept a substrategy with CEU value less by $\varepsilon_0$ than the best substrategy that he knows to be enforceable (at that point of the recursive process).

The preceding procedure can be easily amended to take into account limited willingness to cooperate at $\varepsilon_0$-level. The new (and realistic) feature is that it can now fail, if each of the undominated substrategies generated is rejected by at least one Self.

A limited cooperation procedure might work as follows (on Figure 3). At node 1, $S_1$ is still the tentative substrategy, whereas $S'_1$ and the other dotted line substrategies are now those substrategies which are known to be acceptable by the node 1 Self and all his successors (this means that $V_1(S'_1) \geq V_1(S_1) - \varepsilon_0$, etc...).

Now a substrategy is no longer available at node 0 when one of its branches is missing. For instance, if (U, $S_1$ if $A_1$, $S''_2$ if $A_2$) is the only $(\alpha_i)$-substrategy containing $S_1$ and $S''_2$ is not acceptable at node 2, then "$S_1$ if $A_1$" cannot be considered at node 0. Thus, after some more chopping off has taken place, and, at each node k, $S_k$ has been replaced by the best remaining substrategy for criterion $V_k$, the procedure can continue exactly as in the unlimited case.

It may of course happen that there does not remain any available substrategy at node 0, in which case $\varepsilon_0$-cooperation should be considered to have failed.

## 5  CONCLUSION

Thus, the fact that the resolute choice approach must give up to separability does not necessarily prevent the existence of operational ways of constructing a resolute choice solution. The procedures we have proposed are not the only conceivable ones, and it remains to examine whether or not the solutions they favour reflect more likely compromises between the different selves than alternativeprocedures.

Extensions of the procedures to the construction of strategies which satisfy other types of non-dominance requirements, such as w.r.t. stochastic dominance under risk, only seem to require a suitable adaptation of the strategy generating mecanism.

We hope that the above considerations demonstrate that decision aiding has no theoretical nor practical ground to limit itself to SEU theory.